\documentclass[11pt,a4paper]{article}
\usepackage[T1]{fontenc}
\usepackage{lmodern}
\usepackage[margin=28mm]{geometry}
\usepackage{amsmath,amssymb,amsthm,bm}
\usepackage{graphicx,flafter}
\graphicspath{{drafts/FullESPRIT/figures/}{figures/}}

\usepackage{needspace}
\usepackage{cite}
\usepackage[hidelinks]{hyperref}
\AddToHook{env/thebibliography/begin}{\raggedright}
\numberwithin{equation}{section}
\numberwithin{figure}{section}
\numberwithin{table}{section}
\newtheorem{theorem}{Theorem}[section]
\newtheorem{lemma}[theorem]{Lemma}

\theoremstyle{definition}
\newtheorem{remark}[theorem]{Remark}
\DeclareMathOperator{\md}{md}
\DeclareMathOperator{\diag}{diag}
\DeclareMathOperator{\range}{range}
\newcommand{\T}{\mathbb T}
\title{\mbox{Stability Analysis of Single- and Multi-Snapshot ESPRIT}}
\author{Wenchong Huang\thanks{\raggedright
School of Mathematical Sciences, Zhejiang University, Hangzhou, 310027,
China. (\mbox{vongvuncung@zju.edu.cn}, 
\mbox{pingliu@zju.edu.cn})}
\and Ping Liu\footnotemark[1]\ \thanks{%
Institute of Fundamental and Transdisciplinary Research, Zhejiang
University, Hangzhou, 310027, China.}}
\date{}
\begin{document}
\maketitle

\begin{abstract}
Spectral estimation seeks to recover unknown frequencies from $T$ snapshots, each consisting of $\Omega+1$ noisy Fourier measurements. Super-resolution refers to the recovery of frequencies whose minimum separation $\Delta$ is smaller than $1/\Omega$.
We establish state-of-the-art stability bounds for single- and multi-snapshot ESPRIT when the frequencies form well-separated clumps, with the largest clump containing $n^\star$ frequencies. In particular, for single-snapshot ESPRIT, our error bound scales as 
\[
\mathcal O\bigl(\Omega^{-1}\mathrm{SRF}^{2n^\star-2}
\cdot\mathrm{Noise}\bigr),
\]
where $\mathrm{SRF}:=(\Omega\Delta)^{-1}$ denotes the super-resolution factor, provided that the noise level satisfies
$\mathrm{Noise}\lesssim(\Omega\Delta)^{2n^\star-2}$.
In comparison with the minimax bounds \cite{batenkov2019super} and the computational resolution limits \cite{liu2021mathematical, liu2024mathematical}, our result shows that single-snapshot ESPRIT achieves the optimal stability scaling in the super-resolution regime. For multi-snapshot ESPRIT, we establish error bounds under 
both deterministic noise and Gaussian white noise, 
attaining the best-known SRF dependence under weaker conditions. 
Numerical experiments corroborate the theoretical error bounds.
\end{abstract}


\IfFileExists{drafts/FullESPRIT/sections/introduction.tex}
  {\section{Introduction}
\label{sec:introduction}

The problem of recovering source locations or characteristic frequencies from noisy
measurements arises in many interesting applications in sensing and imaging, including
direction-of-arrival estimation~\cite{Capon1969,Schmidt1986,RK},
inverse source imaging~\cite{MosherLewisLeahy1992,Fannjiang2009,FannjiangStrohmerYan2010},
inverse scattering~\cite{Cheney2001,Kirsch2002,GruberMarengoDevaney2004,AmmariIakovlevaLesselier2005,DevaneyMarengoGruber2005,Fannjiang2010},
nuclear magnetic resonance spectroscopy~\cite{LinEtAl1997}, and Hamiltonian
identification in superconducting quantum processors~\cite{HangleiterEtAl2024}.

The model problem can be formulated as follows. Given the noisy
measurements
\begin{equation}\label{eq:intro-model}
 y_{k,t}=\sum_{j=1}^n a_{j,t}e^{-2\pi i k\omega_j}+\eta_{k,t},
 \qquad k=0,\ldots,\Omega,\quad t=1,\ldots,T,
\end{equation}
where $\mathcal X:=\{\omega_1,\ldots,\omega_n\}\subset
\mathbb T:=\mathbb R/\mathbb Z$ is an unknown set of distinct
frequencies, $a_{j,t}\in\mathbb C$ are amplitudes,
$\eta_{k,t}\in\mathbb C$ is measurement noise, and $T$ is the number of
snapshots, the goal is to recover the frequencies.
The minimum separation and super-resolution factor (SRF)~\cite{LLF} are defined by
\begin{equation}\label{eq:intro-separation-srf}
 \Delta:=\min_{j\ne\ell}|\omega_j-\omega_\ell|_{\mathbb T}>0,
 \qquad \mathrm{SRF}:=(\Omega\Delta)^{-1},
\end{equation}
where $|x|_{\mathbb T}:=\min_{m\in\mathbb Z}|x-m|$.
Common approaches to spectral estimation include
Fourier-based methods~\cite{Bartlett1950,Welch1967}, optimization-based methods~\cite{ZiskindWax1988,MalioutovCetinWillsky2005,CandesFernandezGranda2014,TangEtAl2013,BhaskarTangRecht2013},
Bayesian methods~\cite{Bretthorst1988,AndrieuDoucet1999,WipfRao2004,YangXieZhang2013,BadiuEtAl2017}, and Prony-type methods.
The regime $\mathrm{SRF}\ge1$ is referred to as the super-resolution
regime~\cite{LLF}, where conventional Fourier-based methods may fail to
resolve closely spaced frequencies~\cite{Capon1969,LiaoFannjiang2016}. In contrast, Prony-type methods, including Prony's
method~\cite{Prony1795,TuftsKumaresan1982,BeylkinMonzon2005},
MUSIC~\cite{BienvenuKopp1983,Schmidt1986},
ESPRIT~\cite{RPK,RK},
and the matrix pencil method~\cite{HuaSarkar1990}, can resolve these closely
spaced frequencies provided that the noise is sufficiently
small~\cite{LiaoFannjiang2016,LLF,LZGL}.

Among these methods, MUSIC and ESPRIT are two representative
signal-subspace approaches to super-resolution, 
which exploit the underlying low-rank signal structure.
MUSIC estimates frequencies by finding steering vectors that are nearly
orthogonal to the estimated noise subspace.
It was developed independently by Schmidt~\cite{Schmidt1986} and by
Bienvenu and Kopp~\cite{BienvenuKopp1983} for direction-of-arrival estimation.
Barabell~\cite{Barabell1983} improved its efficiency by
replacing the spectral search with polynomial rooting,
yielding Root-MUSIC.
Further developments include spatial smoothing~\cite{SWK,PillaiKwon1989},
wideband processing~\cite{WangKaveh1985},
Root-MUSIC variants~\cite{Friedlander1993,PesaventoGershmanHaardt2000},
and recursive MUSIC~\cite{MosherLeahy1998,MosherLeahy1999}.
The first statistical analysis of MUSIC was probably given by Kaveh and
Barabell~\cite{KavehBarabell1986}, who derived an approximate resolution
threshold for two uncorrelated sources of equal power.
Later work derived multi-snapshot error bounds for MUSIC with
multiple sources~\cite{StoicaNehorai1989} and for
Root-MUSIC~\cite{RaoHariRootMUSIC1989}.

The first deterministic stability analysis of single-snapshot MUSIC was given by
Liao and Fannjiang~\cite{LiaoFannjiang2016}. They developed a deterministic framework for analyzing the stability of single-snapshot MUSIC through singular-value estimates for Vandermonde matrices. They also established an explicit $\mathcal{O}(\Omega^{-1}\cdot \mathrm{Noise})$ error bound for the noise-space correlation function, provided that the well-separated condition $\mathrm{SRF}<1/2$ holds. This analysis was also generalized to
higher dimensions~\cite{Liao2015}.
The later work~\cite{LiLiao2021} followed this framework and obtained an $\mathcal{O}(\Omega^{-1} \mathrm{SRF}^{2n^\star-2}\cdot \mathrm{Noise})$ error bound under the separated clumps model with $\mathrm{SRF}\geq 1$, where $n^\star$ denotes the largest clump size.
Li et al.~\cite{LZGL} further extended the super-resolution
stability analysis to multi-snapshot MUSIC and ESPRIT.

ESPRIT instead estimates frequencies through
rotational invariance of the signal subspace
and has attracted considerable attention for its exceptional
computational efficiency, numerical stability, and super-resolution capabilities.
It was originally proposed by Paulraj, Roy, and Kailath~\cite{PaulrajRoyKailath1985,RPK},
developed by Roy and Kailath~\cite{RK}, and
later generalized to higher
dimensions~\cite{ZoltowskiHaardtMathews1996,HaardtNossek1998,HaardtRoemerDelGaldo2008}.
Unitary ESPRIT~\cite{HaardtNossek1995} further reduced computational cost through
real-valued computations and improved estimation accuracy,
while structured least squares~\cite{Haardt1997} improved accuracy by
preserving the structure of the shift-invariance equation.
Early analyses quantified its asymptotic performance and sensitivity to
noise~\cite{RaoHari,OVK,LiLiuVaccaro1993}.
Roemer et al.~\cite{RoemerHaardtDelGaldo2014} later developed a first-order
perturbation framework under
zero-mean noise with finite second moments, with a generalization to higher dimensions.
Aubel and B\"olcskei~\cite{AubelBolcskei} and
Fannjiang~\cite{Fannjiang} established deterministic stability analyses
for single-snapshot ESPRIT, using singular-value estimates for
Vandermonde matrices to obtain frequency error bounds for well-separated
frequencies under small noise.

The first deterministic super-resolution analysis for single-snapshot ESPRIT was given by
Li et al.~\cite{LLF} under the separated clumps model.
Li et al.~\cite{LZGL} also gave super-resolution error bounds
of multi-snapshot ESPRIT with small Gaussian white noise and enough snapshots.
Yang~\cite{Yang} used a different approach based on a matrix
perturbation bound to obtain high-probability frequency error bounds in
terms of the smallest singular value of the Vandermonde matrix.
Another outstanding contribution is due to Ding et al.~\cite{DingEtAl},
who obtained a high-probability
$\widetilde{\mathcal O}\bigl(\Omega^{-3/2}(\mathrm{Noise}^2+\mathrm{Noise}^3)\bigr)$
frequency error bound for single-snapshot ESPRIT under high noise
and for large enough $\Omega$, where the exponent of $\Omega$ is optimal.
We will compare these results with ours later in this article.

\subsection{Review of ESPRIT and our contribution}

For the single-snapshot case $T=1$, suppose that 
$\Omega=2M\ge2n$ and $a_{j,1}\ne0$ for $j=1,\ldots,n$.
Define the Hankel operator
\begin{equation}\label{eq:intro-hankel-operator}
 \mathcal H(v):=(v_{p+q})_{0\le p,q\le M},
 \qquad v\in\mathbb C^{2M+1}.
\end{equation}
For $y=(y_{k,1})_{k=0}^{2M}$ and $\eta=(\eta_{k,1})_{k=0}^{2M}$,
the observation matrix satisfies
\begin{equation}\label{eq:hankel}
 \widehat H:=\mathcal H(y)
 =\Phi_M\operatorname{diag}(a_{1,1},\ldots,a_{n,1})\Phi_M^{T}
   +\mathcal H(\eta),
\end{equation}
where $\Phi_N=(e^{-2\pi i k\omega_j})_{0\le k\le N,\,1\le j\le n}$.
Consider the singular value decomposition (SVD)
\[
\widehat H=\left[\widehat U \quad \widehat U_{\perp}\right]\,\Sigma\, \left[\widehat W \quad \widehat W_{\perp}\right]^*,
\]
where $\widehat{U}\in\mathbb{C}^{(M+1)\times n}$ consists of the leading $n$ left singular vectors. 
Deleting its last and first rows gives $\widehat U_{\uparrow}$ and $\widehat U_{\downarrow}$, respectively.
Single-snapshot ESPRIT~\cite{Fannjiang,LLF} computes the estimated frequencies $\widehat{\omega}_1,...,\widehat{\omega}_n$ by
\begin{equation}\label{eq:intro-single-esprit}
 \widehat\Psi
 :=\widehat U_{\uparrow}^{\dagger}\widehat U_{\downarrow},
 \qquad
 \widehat\omega_j:=-\frac{\arg(\widehat z_j)}{2\pi},
 \quad j=1,\ldots,n,
\end{equation}
where $\widehat z_j$ are the eigenvalues of $\widehat\Psi$ and
$^\dagger$ denotes the Moore--Penrose pseudoinverse.

With noiseless measurements, we have $\Sigma=\operatorname{diag}$ $\left(\sigma_1, \ldots, \sigma_n, 0, \ldots, 0\right)$, which yields $\operatorname{Range}(\widehat{H})=\operatorname{Range}(\widehat{U})=\operatorname{Range}(\Phi_M)$. So, there exists an invertible matrix $R \in \mathbb{C}^{n \times n}$ such that $\widehat{U}R=\Phi_M$. Consequently,
\[
R^{-1}\Psi R=\left(\widehat{U}_\uparrow R\right)^{\dagger}\left(\widehat{U}_\downarrow R\right)=(\Phi_M)^\dagger_\uparrow(\Phi_M)_\downarrow=\operatorname{diag}\left(e^{-2\pi i\omega_1}, \cdots, e^{-2\pi i\omega_n}\right):=\Lambda,
\]
which shows the correctness of single-snapshot ESPRIT.

With noisy measurements, we have the noisy relation $\widehat{U}R=\Phi_M+F$. Therefore,
\begin{align*}
R^{-1}\widehat\Psi R
 =&\left(\widehat{U}_\uparrow R\right)^{\dagger}\left(\widehat{U}_\downarrow R\right) = (\Phi_{M-1} + F_{\uparrow})^{\dagger}(\Phi_{M-1} \Lambda + F_{\downarrow})\\
 =& (\Phi_{M-1} + F_{\uparrow})^{\dagger}(\Phi_{M-1} \Lambda + F_{\uparrow}\Lambda - F_{\uparrow}\Lambda + F_{\downarrow})\\
 =& \Lambda  + G,
\end{align*}
where $G:=(\Phi_{M-1}+F_{\uparrow})^\dagger (F_{\downarrow}-F_{\uparrow}\Lambda)$.
This is already enough to estimate the distance between the eigenvalues of $\widehat\Psi$ and those of $\Lambda$ when $G$ is small enough, which motivates our stability proof.

Leveraging the above strategy, we analyze the stability of ESPRIT for supports consisting of separated clumps, as in~\cite{LiLiao2021,BatenkovEtAl2020,LLF,KunisNagel2020,BatenkovEtAl2021,BatenkovGoldman2021,LZGL}. We write $a_{\min}:=\min_j|a_{j,1}|$.
The table below compares the sufficient noise thresholds and matching-error
bounds in Theorem~\ref{thm:main} with existing estimates.
\begingroup
\renewcommand{\arraystretch}{1.5}
\setlength{\arraycolsep}{1.5pt}
\[
\begin{array}{c|c|c}
 \text{Result} & \text{Threshold for }\|\mathcal H(\eta)\|_2/a_{\min}
 & \text{Error Bound}\\ \hline
 \text{\cite[Theorem~3(a)]{LLF}}
 & \mathcal O\bigl(\Omega\,\mathrm{SRF}^{-(2n^\star-2)}\bigr)
 & \mathcal O\bigl(\Omega^{-1}\mathrm{SRF}^{3n^\star-3}\|\mathcal H(\eta)\|_2/a_{\min}\bigr)\\
 \text{\cite[Theorem~3(b)]{LLF}}
 & \mathcal O\bigl(\mathrm{SRF}^{-(4n^\star-3)}\bigr)
 & \mathcal O\bigl(\Omega^{-1}\mathrm{SRF}^{2n^\star-2}\|\mathcal H(\eta)\|_2/a_{\min}\bigr)\\
 \text{Theorem~\ref{thm:main}}
 & \mathcal O\bigl(\Omega\,\mathrm{SRF}^{-(2n^\star-2)}\bigr)
 & \mathcal O\bigl(\Omega^{-1}\mathrm{SRF}^{2n^\star-2}\|\mathcal H(\eta)\|_2/a_{\min}\bigr)
\end{array}
\]
\endgroup
Our bound attains the error order of the small-noise estimate
in~\cite[Theorem~3(b)]{LLF} under only their moderate-noise condition.
The key steps in their moderate-noise analysis are
\[
 \|\widehat\Psi-\Psi\|_2
 \lesssim
 \frac{\|\mathcal H(\eta)\|_2}
 {a_{\min}\sigma_{\min}(\Phi_M)^2}
 \qquad\text{and}\qquad
 \md(\bm\omega,\widehat{\bm\omega})
 \lesssim
 \frac{\sqrt M}{\sigma_{\min}(\Phi_M)}
 \|\widehat\Psi-\Psi\|_2.
\]
The additional factor $\sqrt M/\sigma_{\min}(\Phi_M)$ in the second
step comes from applying the spectral matching theorem
\cite[Chapter~IV, Theorem~3.3]{SS} to the generally nonnormal matrix $\Psi$.
In our analysis, we write $\widehat UR=\Phi_M+F$ and obtain
\[
 \|R^{-1}\widehat\Psi R-\Lambda\|_2
 \le
 \frac{2\|F\|_2}
 {\sigma_{\min}(\Phi_{M-1})-\|F\|_2}.
\]
Since $\Lambda$ is diagonal, the same spectral matching theorem
yields~\eqref{eq:row-error} without
an additional condition-number factor. Bounding $\|F\|_2$ then gives the error order
$\mathcal O\bigl(\Omega^{-1}\mathrm{SRF}^{2n^\star-2}
\cdot\mathrm{Noise}\bigr)$.

From an information-theoretic perspective, two complementary lines of work have established fundamental limits for super-resolution. A series of works by Liu and collaborators~\cite{liu2021mathematical, LiuZhang2022, liu2024mathematical} quantified the minimum separation required for reliable source-number detection and stable localization from noisy measurements, while the minimax theory~\cite{batenkov2019super, liu2023super} characterized the best achievable worst-case recovery accuracy among all estimators. These results provide fundamental benchmarks for assessing the stability of super-resolution algorithms. By comparing our stability bounds with these information-theoretic limits, we show that single-snapshot ESPRIT achieves the optimal stability scaling in the super-resolution regime, under only the necessary noise condition for stable recovery.

 For multiple snapshots, suppose $T\ge n$ and $\Omega=M\ge n$, and write
\begin{equation}\label{eq:intro-ms-factorization}
 \widehat Y=(y_{k,t})_{k,t}=\Phi_M C+E,
 \qquad C=(a_{j,t})_{j,t}\in\mathbb C^{n\times T},
 \qquad E=(\eta_{k,t})_{k,t}.
\end{equation}
Multi-snapshot ESPRIT~\cite{RK,LZGL} instead applies the SVD to $\widehat Y$ and gives
\[
\widehat Y=\left[\widehat U \quad \widehat U_{\perp}\right]\,\Sigma\, \left[\widehat W \quad \widehat W_{\perp}\right]^*,
\]
where $\widehat{U}\in\mathbb{C}^{(M+1)\times n}$ consists of the leading $n$ left singular vectors. Again, applying (\ref{eq:intro-single-esprit}) gives the estimated frequencies.
Some existing works~\cite{LZGL,Yang} apply the SVD to the sample covariance
matrix $T^{-1}\widehat Y\widehat Y^*$, which is equivalent to the direct SVD
of $\widehat Y$ used here.

We extend the perturbation argument to multi-snapshot ESPRIT, obtaining in
Theorem~\ref{thm:ms-main} the following bound:
\[
 \md(\bm\omega,\widehat{\bm\omega})
 =\mathcal O\bigl((T\Omega)^{-1/2}
 \mathrm{SRF}^{n^\star-1}\|E\|_2\bigr)
\]
under the noise condition
\[
 \|E\|_2\lesssim\sqrt{T\Omega}\,
 \mathrm{SRF}^{-(n^\star-1)}.
\]

When the noise entries are i.i.d.\ complex Gaussian with variance $\sigma^2$,
Theorem~\ref{thm:ms-white-high-probability} further yields $T^{-1/2}$ error decay.
We compare our bound with existing high-probability bounds in the table below
under the same clump geometry, with other parameters fixed.
Note that Yang's original bound~\cite[Theorem~15(1)]{Yang}
is of order $\mathcal O\bigl(T^{-1/2}\Omega\,\sigma_{\min}(\Phi_\Omega)^{-3}
\max\{\sigma\|\Phi_\Omega\|_2,\sigma^2\}\bigr)$,
which yields the second row below after applying Lemma~\ref{lem:clumps}.
\begingroup
\renewcommand{\arraystretch}{1.5}
\setlength{\arraycolsep}{2pt}
\thinmuskip=2mu
\medmuskip=2mu
\thickmuskip=3mu
\[
\begin{array}{@{}c|c|c|c@{}}
 & \text{Noise} & \text{Snapshot} & \text{Error bound}\\ \hline
 \text{\cite[Thm.~IV.6(b)]{LZGL}}
 & \sigma\lesssim \sqrt \Omega\,\mathrm{SRF}^{-(n^\star-1)}
 & T\gtrsim n+\sigma^2\Omega^2\mathrm{SRF}^{6n^\star-4}
 & \mathcal O\bigl(T^{-1/2}\mathrm{SRF}^{n^\star-1}\sigma\bigr)\\
 \text{\cite[Thm.~15(1)]{Yang}}
 & \sigma\lesssim \sqrt \Omega
 & T\gtrsim n+\Omega
 & \mathcal O\bigl(T^{-1/2}\mathrm{SRF}^{3n^\star-3}\sigma\bigr)\\
 \text{Theorem~\ref{thm:ms-white-high-probability}}
 & \sigma\lesssim \sqrt \Omega\,\mathrm{SRF}^{-(n^\star-1)}
 & T\gtrsim n+\sigma^2\mathrm{SRF}^{2n^\star-2}
 & \mathcal O\bigl(T^{-1/2}\mathrm{SRF}^{n^\star-1}\sigma\bigr)
\end{array}
\]
\endgroup
Compared with the first row, our bound has the same error order under
a weaker sufficient snapshot condition.
Compared with the second row, our bound improves the SRF power
from $3n^\star-3$ to $n^\star-1$ under the displayed noise condition.
}
  {}

\IfFileExists{drafts/FullESPRIT/sections/single_snapshot.tex}
  {\section{Stability of single-snapshot ESPRIT}
\label{sec:model}

We use the single-snapshot setting of Section~\ref{sec:introduction},
with $\Omega=2M$, and write $a_j=a_{j,1}$.
The matching distance is
\[
 \md(\bm\omega,\widehat{\bm\omega})
 =\min_{\pi\in\mathfrak S_n}\max_j
   |\omega_j-\widehat\omega_{\pi(j)}|_{\T},
\]
where $\mathfrak S_n$ is the set of all permutations of $\{1,\ldots,n\}$.
The following theorem is the main stability result for single-snapshot ESPRIT.

\begin{theorem}\label{thm:main}
Suppose that $\mathcal X$ is partitioned into nonempty
clumps $\mathcal C_1,\ldots,\mathcal C_A$, and let
$n^\star=\max_a|\mathcal C_a|$. Assume that
$M\ge\max\{n,10n^\star\}$, that
$M\Delta\le\tfrac52 n^\star$ whenever $n^\star\ge2$, and that
\begin{equation}\label{eq:geometry}
 \operatorname{dist}_{\T}(\mathcal C_a,\mathcal C_b)
 >\frac{5n^\star}{M}
 \quad(a\ne b).
\end{equation}
If
\begin{equation}\label{eq:noise-explicit}
 \|\mathcal H(\eta)\|_2\le
 \frac{a_{\min}M}{15nn^\star6^{n^\star}}
 \left(\frac{2M\Delta}{5n^\star}\right)^{2n^\star-2},
\end{equation}
then $\widehat U_{\uparrow}$ has full column rank, all $\widehat z_j$ are nonzero,
and
\begin{equation}\label{eq:error-explicit}
 \md(\bm\omega,\widehat{\bm\omega})
 \le
 \frac{16nn^\star(2n-1)6^{n^\star}}{\sqrt{15}}
 \left(\frac{5n^\star}{2M\Delta}\right)^{2n^\star-2}
 \frac{\|\mathcal H(\eta)\|_2}{a_{\min}M}.
\end{equation}
\end{theorem}

We measure accuracy by the root mean square error over $N$
independent trials,
\begin{equation}\label{eq:rmse}
 \mathrm{RMSE}:=
 \left(\frac1N\sum_{\ell=1}^N
 \md(\bm\omega,\widehat{\bm\omega}^{(\ell)})^2\right)^{1/2}.
\end{equation}
Throughout the numerical experiments, we take $N=64$. We test the SRF and noise orders
in Theorem~\ref{thm:main}. For $n^\star\in\{2,3,4\}$, set
\[
 \mathcal C_1=\{0.2+j\Delta:0\le j<n^\star\},\qquad
 \mathcal C_2=\{0.7+j\Delta:0\le j<n^\star\},\qquad
 \mathcal X=\mathcal C_1\cup\mathcal C_2.
\]
We take $a_j=1$ for every source and fix $M=256$.
The results in Figure~\ref{fig:single-scaling} agree closely with
the SRF power $2n^\star-2$ and linear noise dependence in
Theorem~\ref{thm:main}.

\begin{figure}[!htb]
 \centering
 \includegraphics[width=\linewidth]{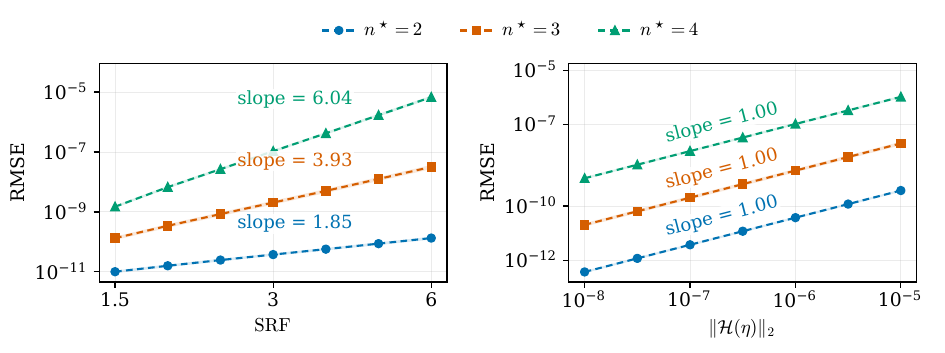}
 \caption{Testing the SRF and noise orders in
 Theorem~\ref{thm:main}. Left: $\|\mathcal H(\eta)\|_2=10^{-6}$,
 with $\Delta$ varied.
 Right: $\mathrm{SRF}=3$, with $\|\mathcal H(\eta)\|_2$ varied.
 Shaded bands are pointwise $95\%$ bootstrap intervals.}
 \label{fig:single-scaling}
\end{figure}

\subsection{Proof of Theorem \ref{thm:main}}
\label{sec:proof}

The retained rows of the left Hankel factor in (\ref{eq:hankel}) form $\Phi_{M-1}$.
The following estimate is adapted and refined from
\cite[Theorem~3.10]{Li}.
\begin{lemma}\label{lem:clumps}
Under the geometric assumptions of Theorem~\ref{thm:main},
\begin{equation}\label{eq:li-bound}
 \sigma_{\min}(\Phi_M)\ge\sigma_{\min}(\Phi_{M-1})
 \ge
 \sqrt{\frac{M}{3nn^\star6^{n^\star}}}
 \left(\frac{2M\Delta}{5n^\star}\right)^{n^\star-1},
\end{equation}
where $\sigma_{\min}$ denotes the smallest nonzero singular value.
\end{lemma}

\Needspace{7\baselineskip}
The singular vectors can be expressed in the original Fourier basis.
The following lemma applies to both the Hankel and measurement matrices.
\begin{lemma}\label{lem:basis}
Let $C\in\mathbb C^{n\times T}$ have full row rank, and let
$\widehat U$ contain the leading $n$ left singular vectors of
$\widehat Y=\Phi_M C+E$. If
\[
 \|E\|_2\le\frac15\sigma_{\min}(\Phi_M)\sigma_{\min}(C^*),
\]
then there exist an invertible $R$ and a matrix $F$ such that
\begin{equation}\label{eq:basis-error}
 \widehat UR=\Phi_M+F,
 \qquad
 \|F\|_2\le\frac{4\|E\|_2}{\sqrt{15}\,\sigma_{\min}(C^*)}.
\end{equation}
\end{lemma}
\begin{proof}
Consider the singular value decomposition of $\widehat Y$,
\[
\widehat Y= [\widehat U \ \widehat U_{\perp}]\diag(\widehat \Sigma, \widehat \Sigma_{res}) [\widehat W \ \widehat W_{\perp}]^*,
\]
where $\widehat W,\widehat\Sigma$ are the right singular
vectors and singular values corresponding to the first $n$ elements. Since $\widehat Y=\Phi_M C+E$, we have 
\[
\widehat U \widehat \Sigma  = \Phi_M C\widehat W+E\widehat W. 
\]
Define $B:=C\widehat W$. We then construct the desired decomposition (\ref{eq:basis-error}),
\[
\widehat UR=\Phi_M+F
\]
for $R:=\widehat\Sigma B^{-1}$ and $F:=E\widehat WB^{-1}$ if $B$ is invertible.

To prove the invertibility of $B$, let $\Pi$ be the orthogonal projection onto
$\range(C^*)$. Then
\[
 \|Cw\|_2\ge\sigma_{\min}(C^*)\|\Pi w\|_2.
\]
Weyl's inequality \cite[Theorem~3.3.16(c)]{HornJohnson1991} gives
\[
 \sigma_n(\widehat Y)\ge
 \sigma_{\min}(\Phi_M)\sigma_{\min}(C^*)-\|E\|_2\geq \frac{4}{5} \sigma_{\min}(\Phi_M)\sigma_{\min}(C^*),
\]
and thus $\widehat \Sigma$ is invertible. The SVD of $\widehat Y^*$ gives 
\[
\widehat{Y}^* \widehat{U}=\widehat{W} \widehat{\Sigma}= C^* \Phi_M^*\widehat{U}+E^*\widehat{U}.
\]
Then $(I-\Pi)(\Phi_M C)^*=0$ implies that
\[
 (I-\Pi)\widehat W
 =(I-\Pi)E^*\widehat U\widehat\Sigma^{-1},
\]
which further gives
\[
 \|(I-\Pi)\widehat W\|_2
 \le\frac{\|E\|_2}{\sigma_n(\widehat Y)}\le\frac14.
\]
Consequently,
\[
 \|\Pi\widehat Wv\|_2^2
 =\|v\|_2^2-\|(I-\Pi)\widehat Wv\|_2^2
 \ge\frac{15}{16}\|v\|_2^2.
\]
The matrix $B:=C\widehat W$ is therefore invertible, with
\[
 \|Bv\|_2=  \|C\Pi\widehat Wv\|_2 \ge\frac{\sqrt{15}}4\sigma_{\min}(C^*)\|v\|_2,
 \qquad
 \|B^{-1}\|_2\le\frac4{\sqrt{15}\,\sigma_{\min}(C^*)}.
\]
Both $B$ and $\widehat\Sigma$ are invertible, so $R$ is invertible.
Since $F:=E\widehat WB^{-1}$, the bound on $B^{-1}$ proves \eqref{eq:basis-error}.
\end{proof}

The one-step shift converts the basis perturbation into a support error.
\begin{lemma}\label{lem:row}
Suppose that $\widehat UR=\Phi_M+F$ with $R$ invertible and
$\|F\|_2\le\sigma_{\min}(\Phi_{M-1})/4$. Then $\widehat U_{\uparrow}$ has full
column rank, all eigenvalues of $\widehat\Psi$ are nonzero, and
\begin{equation}\label{eq:row-error}
 \md(\bm\omega,\widehat{\bm\omega})
 \le\frac{4(2n-1)}{3}
       \frac{\|F\|_2}{\sigma_{\min}(\Phi_{M-1})}.
\end{equation}
\end{lemma}
\begin{proof}
Put
\[
 z_j=e^{-2\pi i\omega_j},\qquad
 \Lambda=\diag(z_1,\ldots,z_n),\qquad
 u=\frac{\|F\|_2}{\sigma_{\min}(\Phi_{M-1})}\le\frac14.
\]
Applying the same row deletions gives
\[
 \widehat U_{\uparrow}R=\Phi_{M-1}+F_{\uparrow},\qquad
 \widehat U_{\downarrow}R=\Phi_{M-1}\Lambda+F_{\downarrow},
\]
and hence
\begin{equation}\label{eq:row-retained-singular-value}
 \sigma_{\min}(\Phi_{M-1}+F_{\uparrow})
 \ge \sigma_{\min}(\Phi_{M-1})-\|F_{\uparrow}\|_2
 \ge (1-u)\sigma_{\min}(\Phi_{M-1})>0.
\end{equation}
Thus $\operatorname{rank}(\widehat U_{\uparrow})
=\operatorname{rank}(\Phi_{M-1}+F_{\uparrow})=n$.
Consequently,
\begin{equation}\label{eq:row-exact-esprit}
\begin{aligned}
 R^{-1}\widehat\Psi R
 &=(\widehat U_{\uparrow}R)^\dagger (\widehat U_{\downarrow}R)\\
 &=(\Phi_{M-1}+F_{\uparrow})^\dagger
   (\Phi_{M-1}\Lambda+F_{\downarrow})\\
 &=(\Phi_{M-1}+F_{\uparrow})^\dagger
   (\Phi_{M-1}\Lambda+F_{\uparrow}\Lambda - F_{\uparrow}\Lambda + F_{\downarrow}) \\
 &=\Lambda+G,\qquad
 G:=(\Phi_{M-1}+F_{\uparrow})^\dagger
       (F_{\downarrow}-F_{\uparrow}\Lambda).
\end{aligned}
\end{equation}
The error term satisfies
\[
 \|G\|_2
 \le \frac{2\|F\|_2}{(1-u)\sigma_{\min}(\Phi_{M-1})}
 =\frac{2u}{1-u}\le\frac23<1.
\]
Since $\|\Lambda v\|_2=\|v\|_2$ and $\|G\|_2<1$,
$\Lambda+G$ is invertible, so all $\widehat z_j$ are nonzero.
Applying the spectral matching theorem
\cite[Chapter~IV, Theorem~3.3]{SS} and passing from eigenvalues to
frequencies gives
\[
 \md(\bm\omega,\widehat{\bm\omega})
 \le\frac{2n-1}{2}\|G\|_2
 \le\frac{(2n-1)u}{1-u}
 \le\frac{4(2n-1)}3u.\qedhere
\]
\end{proof}

\Needspace{12\baselineskip}
\begin{proof}[Proof of Theorem~\ref{thm:main}]
Set $C=\diag(\bm a)\Phi_M^T$ and $E=\mathcal H(\eta)$. Then
\[
 \widehat H=\Phi_M C+E,
 \qquad
 \sigma_{\min}(C^*)\ge a_{\min}\sigma_{\min}(\Phi_M)>0.
\]
Lemma~\ref{lem:clumps} and \eqref{eq:noise-explicit} imply
\[
 \|\mathcal H(\eta)\|_2\le\frac{a_{\min}}5
       \sigma_{\min}(\Phi_{M-1})\sigma_{\min}(\Phi_M)
 \le\frac{a_{\min}}5\sigma_{\min}(\Phi_M)^2.
\]
Applying Lemma~\ref{lem:basis} gives $\widehat UR=\Phi_M+F$ with
\[
 \|F\|_2\le\frac{4\|\mathcal H(\eta)\|_2}
      {\sqrt{15}\,a_{\min}\sigma_{\min}(\Phi_M)}
 \le\frac4{5\sqrt{15}}\sigma_{\min}(\Phi_{M-1})
 <\frac14\sigma_{\min}(\Phi_{M-1}).
\]
Lemma~\ref{lem:row} proves the rank and nonvanishing assertions and yields
\begin{equation}\label{eq:error-singular}
 \md(\bm\omega,\widehat{\bm\omega})
 \le\frac{16(2n-1)}{3\sqrt{15}}
 \frac{\|\mathcal H(\eta)\|_2}
 {a_{\min}\sigma_{\min}(\Phi_{M-1})\sigma_{\min}(\Phi_M)}.
\end{equation}
Substituting \eqref{eq:li-bound} in \eqref{eq:error-singular} gives
\eqref{eq:error-explicit}.
\end{proof}
}
  {}

\IfFileExists{drafts/FullESPRIT/sections/multi_snapshot.tex}
  {\section{Multi-snapshot ESPRIT}
\label{sec:multisnapshot}

We use the multi-snapshot setting of Section~\ref{sec:introduction},
with $\Omega=M$, and define the empirical amplitude covariance by
\[
 R_a:=\frac1T CC^*.
\]
The following theorem is the main stability result for multi-snapshot ESPRIT.

\begin{theorem}\label{thm:ms-main}
Assume the clump geometry and sampling conditions of
Theorem~\ref{thm:main}. Let $T\ge n$, and suppose that
\eqref{eq:intro-ms-factorization} holds with
\begin{equation}\label{eq:ms-covariance}
 \lambda_{\min}(R_a)\ge\rho>0.
\end{equation}
If
\begin{equation}\label{eq:ms-noise-explicit}
 \|E\|_2\le
 \frac{\sqrt{TM\rho}}{5\sqrt{3nn^\star6^{n^\star}}}
 \left(\frac{2M\Delta}{5n^\star}\right)^{n^\star-1},
\end{equation}
then the retained matrix $\widehat U_{\uparrow}$ has full column rank,
all computed roots are nonzero, and
\begin{equation}\label{eq:ms-error-explicit}
 \md(\bm\omega,\widehat{\bm\omega})
 \le\frac{16(2n-1)\sqrt{nn^\star6^{n^\star}}}{3\sqrt5}
 \left(\frac{5n^\star}{2M\Delta}\right)^{n^\star-1}
 \frac{\|E\|_2}{\sqrt{TM\rho}}.
\end{equation}
The conclusion holds for every choice of leading singular vectors and
every complete eigenvalue enumeration.
\end{theorem}
\begin{proof}
By \eqref{eq:ms-covariance},
\[
 \|C^*v\|_2^2=T v^*R_av\ge T\rho\|v\|_2^2,
 \qquad \sigma_{\min}(C^*)\ge\sqrt{T\rho}>0.
\]
Lemma~\ref{lem:clumps} and \eqref{eq:ms-noise-explicit} imply
\[
 \|E\|_2\le\frac{\sqrt{T\rho}}5\sigma_{\min}(\Phi_{M-1})
 \le\frac{\sqrt{T\rho}}5\sigma_{\min}(\Phi_M).
\]
Applying Lemma~\ref{lem:basis} gives
\[
 \widehat UR=\Phi_M+F,
 \qquad
 \|F\|_2\le\frac{4\|E\|_2}{\sqrt{15}\sqrt{T\rho}}
 \le\frac4{5\sqrt{15}}\sigma_{\min}(\Phi_{M-1})
 <\frac14\sigma_{\min}(\Phi_{M-1}).
\]
Lemma~\ref{lem:row} proves the rank and nonvanishing assertions and gives
\begin{equation}\label{eq:ms-error-singular}
 \md(\bm\omega,\widehat{\bm\omega})
 \le\frac{16(2n-1)}{3\sqrt{15}}
 \frac{\|E\|_2}{\sqrt{T\rho}\,\sigma_{\min}(\Phi_{M-1})}.
\end{equation}
Substituting \eqref{eq:li-bound} into \eqref{eq:ms-error-singular}
proves \eqref{eq:ms-error-explicit}.
\end{proof}

\begin{remark}\label{rem:ms-gaussian-amplitudes}
Condition~\eqref{eq:ms-covariance} is reasonable in
multi-snapshot spectral estimation. For example, suppose that the
amplitude vectors are independent across snapshots and
\begin{equation}\label{eq:ms-gaussian-model}
 \bm a_t\sim\mathcal N_{\mathbb C}(0,\Sigma),
 \qquad \lambda_{\min}(\Sigma)>0,
 \qquad C=[\bm a_1,\ldots,\bm a_T].
\end{equation}
For every $0<\delta<1$, if $T\gtrsim n+\log(1/\delta)$, then
$\lambda_{\min}(R_a)\ge\lambda_{\min}(\Sigma)/4$
with probability at least $1-\delta$;
see \cite[Appendix~F, (90)]{Yang}.
Another example is the deterministic model
$a_{j,t}=b_j e^{2\pi i\nu_jt}$, yielding
$\lambda_{\min}(R_a)\ge\tfrac12\min_j|b_j|^2$
for sufficiently large $T$; see \cite[Eq.~(32)]{AubelBolcskei2019} for details.
\end{remark}

We test the SRF and noise orders in
Theorem~\ref{thm:ms-main}, using the same support $\mathcal X$ as in
the single-snapshot experiment. We fix $M=512$ and $T=10^4$.
Following \cite[Section~VII]{LZGL}, we use the amplitude model
\eqref{eq:ms-gaussian-model} with
\begin{equation}\label{eq:ms-experiment-covariance}
 \Sigma=(1-r)I_n+r\mathbf1_n\mathbf1_n^*,
 \qquad r=\tfrac12.
\end{equation}
We rescale a Gaussian noise matrix to prescribe $\|E\|_2$.
The results in Figure~\ref{fig:multi-correlated-scaling} agree closely with the SRF power
$n^\star-1$ and linear noise dependence in
Theorem~\ref{thm:ms-main}.

\begin{figure}[!htb]
 \centering
 \includegraphics[width=0.9\linewidth]{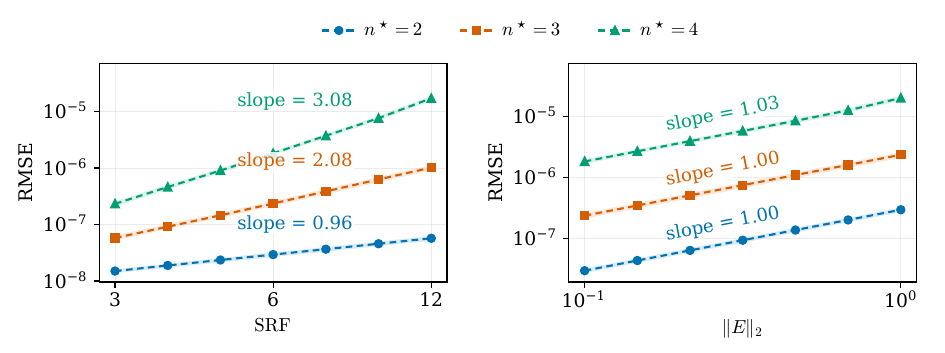}
 \caption{Testing the SRF and noise orders in
 Theorem~\ref{thm:ms-main} with correlated Gaussian amplitudes
 ($r=1/2$). Left: $\|E\|_2=0.1$, with $\Delta$ varied.
 Right: $\mathrm{SRF}=6$, with $\|E\|_2$ varied. Shaded bands are pointwise $95\%$ bootstrap intervals.}
 \label{fig:multi-correlated-scaling}
\end{figure}

\subsection{Gaussian White Noise}
\label{subsec:gaussian-white-noise}

If $|\eta_{k,t}|\le\sigma$ uniformly, a trivial bound is
$\|E\|_2\le\sigma\sqrt{(M+1)T}$.
Under its noise and snapshot conditions,
Theorem~\ref{thm:ms-main} gives an error bound of order
\[
 \frac{\sigma}{\sqrt{\rho}}\,
 \mathrm{SRF}^{n^\star-1}.
\]
However, this bound does not tend to zero as $T$ increases.
In fact, for a sufficiently small fixed $0<h<\Delta/2$, replacing the noiseless
measurements generated by $\mathcal X$ with those generated by $\mathcal X+h$, yields a perturbation satisfying the noise condition
of Theorem~\ref{thm:ms-main}. ESPRIT then recovers $\mathcal X+h$,
so its matching error remains $h$ rather than tending to zero.

However, noise in the real world is usually random. As in~\cite{Yang}, we assume that
\begin{equation}\label{eq:ms-white-noise-model}
 \eta_{k,t}\overset{\mathrm{i.i.d.}}{\sim}
 \mathcal N_{\mathbb C}(0,\sigma^2),\qquad \sigma>0,
\end{equation}
and the noise matrix $E$ is independent of $C$.
Under these assumptions, we can expect the estimation error to tend to zero
as the number of snapshots increases. The following theorem states this fact,
where the proof is given in Appendix~\ref{app:ms-white-proof}.

\begin{theorem}\label{thm:ms-white-high-probability}
Let $M\ge n\ge2$, let the frequencies be distinct in $\mathbb T$, and fix
an amplitude matrix $C$ satisfying \eqref{eq:ms-covariance}. Assume
\eqref{eq:intro-ms-factorization} and \eqref{eq:ms-white-noise-model}.
Suppose that
\begin{equation}\label{eq:ms-white-effective-snr}
 \sigma^2\le\rho\,
                    \sigma_{\min}(\Phi_M)^2.
\end{equation}
For every $0<\delta<1$, suppose that
\begin{equation}\label{eq:ms-white-snapshot-threshold}
 T\ge T_0:=c_0\left[
 n+\log\frac{3}{\delta}
 +\frac{n\sigma^2\bigl(M+1+\log(3/\delta)\bigr)}
 {\rho\sigma_{\min}(\Phi_{M-1})^2}
 \right],
\end{equation}
where $c_0>0$ is an absolute constant.
Then, with probability at least $1-\delta$ over the noise, $\widehat U_\uparrow$ has
full column rank, all computed roots are nonzero, and
\begin{equation}\label{eq:ms-white-sharp}
 \md(\bm\omega,\widehat{\bm\omega})
 \le
 \frac{C(2n-1)\sigma}
 {\sigma_{\min}(\Phi_{M-1})\sqrt{\rho}}\sqrt{\frac{n(M+1+\log(3/\delta))}{T}},
\end{equation}
where $C>0$ is an absolute constant.
\end{theorem}

We test the $T^{-1/2}$ dependence in \eqref{eq:ms-white-sharp} using the support $\mathcal X$ from
the single-snapshot experiment with $n^\star\in\{2,3,4\}$.
We use the source covariance in \eqref{eq:ms-experiment-covariance}
and fix $M=128$, $\mathrm{SRF}=3$, and $\sigma=0.1$.
The slopes in Figure~\ref{fig:ms-white-snapshots} agree closely with the
theoretical order.

\enlargethispage{\baselineskip}
\begin{figure}[!htb]
 \centering
 \includegraphics[width=0.52\linewidth]{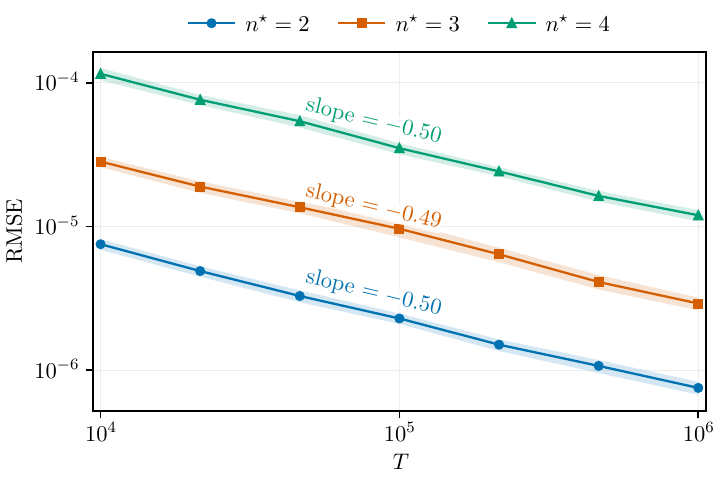}
 \caption{RMSE versus the snapshot count under Gaussian white noise.
 Shaded bands are pointwise $95\%$ paired-bootstrap intervals.
 }
 \label{fig:ms-white-snapshots}
\end{figure}
}
  {}

\IfFileExists{drafts/FullESPRIT/sections/conclusion.tex}
  {\section{Conclusion}
\label{sec:conclusion}

In this paper, we proposed an analysis framework for ESPRIT based on the representation
$\widehat UR=\Phi_M+F$, where $R$ is invertible and $F$ is a small
perturbation. Under this framework, we first gave a concise stability
analysis for single-snapshot ESPRIT, obtaining an error bound of order
$\mathcal O\bigl(\Omega^{-1}\mathrm{SRF}^{2n^\star-2}
\cdot\mathrm{Noise}\bigr)$.
We further extended this analysis to multi-snapshot ESPRIT, obtaining
the error order
$\mathcal O\bigl((T\Omega)^{-1/2}\mathrm{SRF}^{n^\star-1}\|E\|_2\bigr)$.
Exploiting the structure of Gaussian white noise, we then established
in Theorem~\ref{thm:ms-white-high-probability} a high-probability error bound of order
$\mathcal O\bigl(\sigma T^{-1/2}\mathrm{SRF}^{n^\star-1}\bigr)$
under the sufficient snapshot condition
$T\gtrsim n+\sigma^2\mathrm{SRF}^{2n^\star-2}$.
}
  {}

\section*{Acknowledgments}
This work was partially supported by the Fundamental and Interdisciplinary Disciplines Breakthrough Plan of the Ministry of Education 
of China grant number JYB2025XDXM103 and the National Key R\&D Program of China grant number 2024YFA1016000.

\section*{Data Availability Statement}
Data and codes supporting the findings of this work are available upon request.

\section*{Conflict of interest} 
The authors have no conflicts of interest to declare.

\section*{Use of generative AI}
GPT-6 Astra and GPT-6.1 Sol were used to suggest a possible proof strategy for Theorem \ref{thm:ms-white-high-probability} in Appendix A, building on our factorization strategy $\widehat U R = \Phi_M + F$. The resulting Gaussian white-noise analysis is included both for completeness and as an additional finding enabled by our framework. The authors critically examined the AI-generated suggestions, independently verified the mathematical arguments, and take full responsibility for all results and statements in the manuscript.


\par\vspace{\baselineskip}
\appendix
\IfFileExists{drafts/FullESPRIT/sections/gaussian_white_noise_proof.tex}
  {\section{Proof of Theorem~\ref{thm:ms-white-high-probability}}
\label{app:ms-white-proof}

\begin{proof}[Proof of Theorem~\ref{thm:ms-white-high-probability}]
We project $\Phi_M$ onto the estimated signal subspace and put
\[
 P=\widehat U\widehat U^*,\qquad
 R=\widehat U^*\Phi_M,\qquad F=(P-I)\Phi_M.
\]
Then
\[
 \widehat UR=P\Phi_M=\Phi_M+F.
\]
To apply Lemma~\ref{lem:row}, we need to bound $F$ and verify that $R$
is invertible. The norm of $F$ is estimated in
Lemma~\ref{lem:ms-white-projection-residual}: with probability at least $1-\delta$,
\[
 \|F\|_2^2\le\|F\|_{\mathrm F}^2
 \le\frac{c_4n\sigma^2\bigl(M+1+\log(3/\delta)\bigr)}{\rho T}
 \le\frac1{16}\sigma_{\min}(\Phi_{M-1})^2,
\]
where the last inequality holds when $c_0\ge16c_4$. Consequently,
\[
 \sigma_{\min}(\Phi_M+F)
 \ge\sigma_{\min}(\Phi_M)-\|F\|_2
 \ge\frac34\sigma_{\min}(\Phi_M)>0.
\]
Since $\Phi_M+F=\widehat UR$, this proves that $R$ is invertible.
Lemma~\ref{lem:row} now gives full column rank of $\widehat U_\uparrow$,
nonzero computed roots, and
\[
 \md(\bm\omega,\widehat{\bm\omega})
 \le\frac{4(2n-1)}3\frac{\|F\|_2}{\sigma_{\min}(\Phi_{M-1})}
 \le\frac{4\sqrt{c_4}(2n-1)\sigma}
 {3\sigma_{\min}(\Phi_{M-1})\sqrt{\rho}}
 \sqrt{\frac{n\bigl(M+1+\log(3/\delta)\bigr)}T},
\]
where the failure probability is at most $\delta$.
\end{proof}

\subsection{Technical lemmas}
\label{app:ms-white-technical-lemmas}

We will use the following result in random matrix theory; see
\cite[Example~6.2]{Wainwright2019}.

\begin{lemma}\label{lem:ms-white-gaussian-matrix-bounds}
Let $E\in\mathbb{C}^{p\times q}$ have i.i.d.
$\mathcal N_{\mathbb C}(0,1)$ entries, and let $u>0$.

\vspace{0.5em}
\noindent (i) With probability at least $1-e^{-u^2/2}$, we have
\begin{equation}
\label{equ:ms-white-gaussian-matrix-bounds-i}
 \left\|E\right\|_2\le\sqrt p+\sqrt q+u.
\end{equation}
\noindent (ii) With probability at least $1-2e^{-qu^2/2}$, we have
\begin{equation}
\label{equ:ms-white-gaussian-matrix-bounds-ii}
 \left\|q^{-1}EE^*-I\right\|_2\le2\varepsilon+\varepsilon^2,
 \qquad \varepsilon=\sqrt{\frac{p}{q}}+u.
\end{equation}
\end{lemma}

\begin{lemma}\label{lem:ms-white-projection-residual}
Assume the hypotheses of Theorem~\ref{thm:ms-white-high-probability}, and
put $P=\widehat U\widehat U^*$.
There is an absolute constant $c_4>0$ such that, with probability at least
$1-\delta$,
\[
 \|(I-P)\Phi_M\|_{\mathrm F}^2
 \le\frac{c_4n\sigma^2\bigl(M+1+\log(3/\delta)\bigr)}{\rho T}.
\]
\end{lemma}
\begin{proof}
Putting $\Pi=\Phi_M\Phi_M^\dagger$, we obtain from the optimal low-rank
approximation property of the SVD \cite[Theorem~5.9]{TrefethenBau1997} that
\begin{equation}\label{eq:ms-white-svd-residual}
 \|(I-P)\widehat Y\|_{\mathrm F}^2
 \le\|(I-\Pi)\widehat Y\|_{\mathrm F}^2
 =\|(I-\Pi)E\|_{\mathrm F}^2.
\end{equation}
By \eqref{eq:ms-covariance}, $R_a=CC^*/T$ is positive definite. Put
\[
 z=\|(I-P)\Phi_MR_a^{1/2}\|_{\mathrm F}
   =\frac1{\sqrt T}\|(I-P)\Phi_MC\|_{\mathrm F}.
\]
Both projections have rank $n$, so $\operatorname{tr}(P-\Pi)=0$.
Substituting $\widehat Y=\Phi_MC+E$ into \eqref{eq:ms-white-svd-residual}
and expanding the left-hand side gives
\[
 Tz^2+2\Re\operatorname{tr}\left(E^*(I-P)\Phi_MC\right)
 +\|(I-P)E\|_{\mathrm F}^2
 \le\|(I-\Pi)E\|_{\mathrm F}^2,
\]
where we used $(I-P)^*(I-P)=I-P$ and the definition of $z$.
Since $P$ and $\Pi$ are orthogonal projections,
\[
 \|(I-\Pi)E\|_{\mathrm F}^2-\|(I-P)E\|_{\mathrm F}^2
 =\operatorname{tr}\left(E^*(P-\Pi)E\right)
 =\operatorname{tr}\left(EE^*(P-\Pi)\right).
\]
Rearranging and dividing by $T$, we obtain
\[
 \begin{aligned}
 z^2
 &\le-\frac2T\Re\operatorname{tr}\left(E^*(I-P)\Phi_MC\right)
       +\frac1T\operatorname{tr}\left(EE^*(P-\Pi)\right)\\
 &=-2\Re\operatorname{tr}\left(W^*(I-P)\Phi_MR_a^{1/2}\right)
       +\operatorname{tr}\left(N(P-\Pi)\right)\\
 &\le2z\|W\|_{\mathrm F}
       +\left|\operatorname{tr}\left(N(P-\Pi)\right)\right|,
 \end{aligned}
\]
where $W=T^{-1}EC^*R_a^{-1/2}$ and $N=T^{-1}EE^*-\sigma^2I$.
The last inequality follows from Cauchy--Schwarz for the Frobenius inner product.
For the trace term, we have
\[
 \left|\operatorname{tr}\left(N(P-\Pi)\right)\right|
 \le\sum_j\sigma_j(N)\sigma_j(P-\Pi)
 \le\sqrt{2n}\|N\|_2\|P-\Pi\|_{\mathrm F}
 =2\sqrt n\|N\|_2\|(I-P)\Pi\|_{\mathrm F},
\]
where the first inequality is von Neumann's trace inequality
\cite[Eq.~(3.3.25), p.~183]{HornJohnson1991}, the second follows from
Cauchy--Schwarz and $\operatorname{rank}(P-\Pi)\le2n$, and the last equality uses
$\|P-\Pi\|_{\mathrm F}^2=2\|(I-P)\Pi\|_{\mathrm F}^2$.
Moreover, $\Pi=\Phi_M\Phi_M^\dagger$ and the definition of $z$ imply
\[
 \|(I-P)\Pi\|_{\mathrm F}
 \le\frac{\|(I-P)\Phi_M\|_{\mathrm F}}{\sigma_{\min}(\Phi_M)}
 \le\frac{z}{\sqrt{\lambda_{\min}(R_a)}\,\sigma_{\min}(\Phi_M)}.
\]
Substituting these estimates gives 
\[
 z^2
 \le 2z\|W\|_{\mathrm F} + \frac{2 z\sqrt n}{\sqrt{\lambda_{\min}(R_a)}\sigma_{\min}(\Phi_M)}\|N\|_2.
\]
Then dividing by $z$ yields
\begin{equation}\label{eq:ms-white-projection-perturbation}
 \|(I-P)\Phi_M\|_{\mathrm F}
 \le\frac2{\sqrt{\lambda_{\min}(R_a)}}\|W\|_{\mathrm F}
 +\frac{2\sqrt n}{\lambda_{\min}(R_a)\sigma_{\min}(\Phi_M)}\|N\|_2.
\end{equation}
If $z=0$, the residual vanishes and the same bound holds.

It remains to estimate $\|W\|_{\mathrm F}$ and $\|N\|_2$, since
\eqref{eq:ms-covariance} gives $\lambda_{\min}(R_a)\ge\rho$ deterministically.
Put $t=\log(3/\delta)$. For the fixed matrix $C$, the identity
\[
 (C^*R_a^{-1/2})^*(C^*R_a^{-1/2})=T I_n
\]
and the Gaussian noise model imply that $W$ has i.i.d.\
$\mathcal N_{\mathbb C}(0,\sigma^2/T)$ entries.
Applying Lemma~\ref{lem:ms-white-gaussian-matrix-bounds}(i) with
$u=\sqrt{2t}$ gives, with failure
probability at most $e^{-t}$,
\[
 \|W\|_{\mathrm F}\le\sqrt n\|W\|_2
 \le\frac{\sigma\sqrt n}{\sqrt T}
        \left(\sqrt{M+1}+\sqrt n+\sqrt{2t}\right)
 \le c_5\sigma\sqrt{\frac{n(M+1+t)}T}.
\]
Lemma~\ref{lem:ms-white-gaussian-matrix-bounds}(ii), applied to $E/\sigma$
with $u=\sqrt{2t/T}$, gives
\[
 \|N\|_2\le c_6\sigma^2
 \left(\sqrt{\frac{M+1+t}T}+\frac{M+1+t}T\right)
\]
with failure probability at most $2e^{-t}$.
Combining these two estimates, with probability at least $1-3e^{-t}=1-\delta$,
\eqref{eq:ms-white-projection-perturbation} yields
\[
 \|(I-P)\Phi_M\|_{\mathrm F}
 \le c_7\sigma\sqrt{\frac{n(M+1+t)}{\rho T}}
 \left[1+\frac{\sigma}{\sqrt{\rho}\sigma_{\min}(\Phi_M)}
              \left(1+\sqrt{\frac{M+1+t}T}\right)\right].
\]
The effective signal-to-noise and snapshot conditions give, respectively,
\[
 \frac{\sigma}{\sqrt{\rho}\sigma_{\min}(\Phi_M)}\le1,
 \qquad
 \frac{\sigma}{\sqrt{\rho}\sigma_{\min}(\Phi_M)}
       \sqrt{\frac{M+1+t}T}\le\frac1{\sqrt{c_0n}}.
\]
Thus the bracket is bounded by an absolute constant. Squaring proves the
claimed estimate.
\end{proof}
}
  {}

\clearpage
\bibliographystyle{unsrt}
\IfFileExists{drafts/FullESPRIT/references.bib}
  {\bibliography{drafts/FullESPRIT/references}}
  {\bibliography{references}}
\end{document}